\documentclass[letterpaper]{llncs}

\usepackage{epsfig,endnotes}
\usepackage[usenames,dvipsnames]{color}
\usepackage[english,plain]{fancyref}
\usepackage{times}
\usepackage{subfigure}
\usepackage{multirow}
\usepackage{amsfonts,xspace}
\usepackage{url}
\usepackage{graphicx}
\usepackage[sort]{cite}
\usepackage{verbatim}

\newcommand{\ie}{{\it i.e.}\xspace}
\newcommand{\eg}{{\it e.g.}\xspace}

\newcommand{\postfigspace}{-1em}

\newenvironment{packeditemize}{\begin{list}{$\bullet$}{\setlength{\itemsep}{0.2pt}\addtolength{\labelwidth}{-4pt}\setlength{\leftmargin}{\labelwidth}\setlength{\listparindent}{\parindent}\setlength{\parsep}{1pt}\setlength{\topsep}{0pt}}}{\end{list}}

\title{Client-Side Web Proxy Detection \\
from Unprivileged Mobile Devices } 
\author{Huijing Zhang \ \ \ \ David Choffnes \\
\{jessica, choffnes\}@ccs.neu.edu}
\institute{Northeastern University}
\date{\now}
\begin{document}
\maketitle

\section*{Abstract}
Mobile devices that connect to the Internet via cellular networks are rapidly becoming the primary medium for accessing Web content. 
Cellular service providers (CSPs) commonly deploy Web proxies and other middleboxes for security, performance optimization and traffic engineering reasons. 
However, the prevalence and policies of these Web proxies are generally opaque to users and difficult to measure without privileged access to devices and servers. 
In this paper, we present a methodology to detect the presence of Web proxies without requiring access to low-level packet traces on a device, nor access to servers being contacted. 
We demonstrate the viability of this technique using controlled experiments, and present the results of running our approach on several production networks and popular Web sites. 
Next, we characterize the behaviors of these Web proxies, including caching, redirecting, and content rewriting. 
Our analysis can identify how Web proxies impact network performance, and inform policies for future deployments. Last, we release an Android app called Proxy Detector on the Google Play Store, allowing average users with unprivileged (non-rooted) devices to understand Web proxy deployments and contribute to our IRB-approved study. We report on results of using this app on 11 popular carriers from the US, Canada, Austria, and China. 

\vspace{-1em}
\section{Introduction}

As mobile devices such as smartphones and tablets become increasingly ubiquitous, cellular data networks 
have expanded to serve their rising network traffic demands. In part due to scarce and costly bandwidth, 
cellular service providers (CSPs) deploy Web proxies and other middleboxes to efficiently use these scarce 
resources, and to enhance network performance. 

Several prior studies identify and characterize Web and performance-enhancing proxies (PEPs) in cellular 
networks~\cite{ivanovich2008tcp, wang:middleboxes, farkas2012split, pam14, xu-transparent-proxies,gomez2008web,meyer2003performance, wei2006inference, shelby2001performance, necker2005performance}.  These features include transparently proxying TCP connections, 
caching content, rewriting content, and redirecting traffic. While these policies may help or hurt performance, 
ultimately each carrier's proxy behaviors are opaque, meaning that researchers and users have little to 
no knowledge about their impact on network performance for the applications they use. Further, there 
is recent evidence that these behaviors may not be evenly applied to all traffic~\cite{xu-transparent-proxies}.

The goal of our work is to to shed light on these Web proxies for arbitrary networks, and understand their impact 
on network performance for arbitrary destinations. In this paper, we are the first to develop measurement 
techniques that allow us to reliably infer whether there is a proxy in the network, and what 
are its features, \emph{without control of the servers contacted, and without requiring access 
to TCP/IP packet headers, which typically requires root privileges on devices}. We then use this technique to understand the performance 
impact of these proxies on Web object fetch times for some of the most popular Web sites, 
for 11 carriers. 

Our inference technique is based on the observation that the Web proxies we observed in 
mobile networks will interpose on traffic to port 80 (HTTP) but will not interpose on port 443 (HTTPS), 
presumably because they assume that HTTPS encryption will prevent many proxy features 
from working (\eg, caching, transcoding and redirection). We thus identify servers that serve both 
HTTP and HTTPS traffic, and use tests on port 443 as our ``control'' that is not subject to proxy 
policies. 

We infer the existence of Web proxies using differences in handshake latency and object fetch time when comparing control traffic 
and potentially exposed traffic. A key challenge is how to reliably use latency differences to identify 
proxy behavior in the presence of significant noise in performance received by mobile devices in 
cellular networks. We discuss how we develop techniques to achieve statistically sound results 
based on observed network performance and its variance.
In addition to the latency-based inferences, we include tests that build upon previous work 
for detecting content modification and traffic redirection. 
We validate all of our techniques with tests using mobile devices and servers where we have full 
control.


\noindent\textbf{Key Results.}
Our key results are as follows:
\begin{packeditemize}
\item We characterize Web proxy deployments in 11 carriers and find differences even from a study conducted one year ago. 
Those differences indicate proxy features are diminishing in use and change over time. During our study, one carrier (Black Wireless) 
went from having no proxy to having one that compresses CSS and JavaScript. Additionally, T-Mobile no longer 
uses DNS redirection as found in a study from last year.
\item We find transcoding is not common, but even carriers using the same underlying cellular infrastructure do transcoding differently.  
\item Even when carriers use caching, it is used in limited ways, affecting only CSS, JavaScript, GIF, JPG, and GIF files.
\item Web proxies can significantly improve end-to-end performance by reducing object fetch time, but the impact may also be small and 
depends on the content and destination server.
\item All the MVNOs we measured except Black Wireless share the same Web proxy features with their host network. Black Wireless deploys a compression proxy which modifies CSS and JavaScript files, but its host network AT\&T does not do so.
\end{packeditemize}

We have implemented an Android app that implements our approach and deployed it on the Google Play Store. By 
revealing Web proxy deployments and their impact on performance worldwide, we hope to inform future 
implementation decisions for operators, help users understand the policies in their providers, and reveal 
ISP practices for policymakers and regulators.
 

\vspace{-1em}
\section{Related work}

Performance-enhancing proxies (PEPs) have been the focus of a large number of research efforts. These 
devices can improve performance by PEPs and other middleboxes that have been studied extensively in previous work, including work that identifies NAT behaviors~\cite{wang:middleboxes}, as well as split TCP connections~\cite{ivanovich2008tcp}, compression, caching, and transcoding~\cite{netalyzr,xu-transparent-proxies}.

A number of studies focus on the problem of detecting proxies. Rodriguez et al.~\cite{rodriguez2004performance} investigate compressing proxies, using experiments where they control both the client and server being contacted. Similarly, the Netalyzr project~\cite{netalyzr} investigated Web proxies using a Web browser running a Java applet, and identified caching and content rewriting. In follow-up work, Weaver et al.~\cite{pam14} used several new techniques to identify Web proxies and identified cases of proxy redirection.  Both of these studies require a client to run a Java-based applet (which is generally not supported on smartphones), and control of the destination server for measurement traffic. In recent work, Vallina-Rodrigues et al.~\cite{vallina2015beyond} extend this approach to an 
Android app; however, the techniques still require control over the server to identify proxies.

The TraceBox~\cite{detal2013revealing} project identifies middleboxes (including proxies) by sending probes with increasing TTL values and waiting for ICMP time-exceeded replies. By analyzing the returned quoted packet, TraceBox can detect various packet modifications performed by middleboxes. While effective in some networks, this requires root access to inspect packet headers at clients (generally not available on unmodified mobile OSes) and requires the network to forward ICMP probes (which we find does not happen in many mobile networks).

Our work is inspired by Xu et al.~\cite{xu-transparent-proxies}, who develop techniques for detecting PEPs in major US carriers. We investigate the same proxy features, but do so without requiring root access to a mobile device (for accessing TCP/IP packet headers), nor control over the Web servers that our measurement clients contact. As a result, our techniques are suitable for crowdsourcing to arbitrary user devices, and can be deployed as part of unprivileged measurement systems such as Mobilyzer~\cite{nikravesh2015mobilyzer} to understand proxy deployments across multiple carriers and over time.

\begin{table*}[t]
\centering
    \begin{tabular}{l @{\hspace{0.8em}}|@{\hspace{0.8em}} c @{\hspace{1em}} c @{\hspace{1em}} c @{\hspace{1em}} c}  
   \setlength{\tabcolsep}{12pt}
 
   & Proxy Exists? & Caching & Rewriting & Redirection  \\ \hline
    AT\&T & \checkmark &  &  &   \\ 
    {\hspace{0.8em}}Black Wireless & \checkmark &  & \checkmark & \\ 
    T-Mobile &  & &  &   \\ 
    {\hspace{0.8em}}Simple Mobile &  &  &  & \\  
    Sprint & \checkmark & \checkmark & \checkmark &   \\ 
    {\hspace{0.8em}}Boost & \checkmark  & \checkmark  & \checkmark  &   \\ 
    Verizon & \checkmark &  &  &  \\ 
    Rogers Wireless &  &  &  & \\ 
    3 (Hutchison) &  &  &  & \\ 
    China Mobile &  &  &  &   \\ 
    China Unicom &  &  &  &   \\ 
    \end{tabular}  
\caption{\textbf{Web proxy deployments observed in our study.} Several popular carriers in the US---AT\&T, Verizon, Sprint, Black Wireless, and Boost---use proxies. However, we found no evidence of proxies for the carriers we measured in Canada, Austria, and China . We also explore how MVNO Web proxy deployments compare with their host network (MVNOs are indented below their host network). We looked at Black Wireless (uses AT\&T), Simple Wireless (uses T-Mobile), and Boost (uses Sprint). All MVNOs we measured except Black Wireless support the same Web proxy features as their host network. Black Wireless deploys a compression proxy while AT\&T does not.}
\label{tab:summaryresults}
\vspace{-2em}
\end{table*}
\vspace{-1em}
\section{Dataset and Methodology}
In this section, we describe the dataset we use to develop and validate our proxy detection approach, as well as to 
understand proxy deployments in the wild. Our study has the following goals. First, we want to determine if a Web proxy exists in a cellular network and how it affects different destinations. Second, we want to enable these measurements from unprivileged devices, allowing average users to help characterize proxy deployments worldwide. Last, we want to understand the impact of these proxies on client-perceived performance.

\noindent\textbf{Dataset.} 
To validate our approach and to characterize US networks, we collected data using controlled experiments from 
Sprint, Verizon, T-Mobile, AT\&T, Boost, Simple Wireless, and Black Wireless.
This dataset was collected between July, 2015 and October, 2015, and comprises more than two million measurement 
samples, from which we selected 500 complete set of measurement to present here. 
The destinations of our traffic are top Web sites according to Alexa~\cite{topweb}, in addition to a server under our control 
running in EC2, which we use for validation of our methodology.

In addition, we released an Android app that implements our approach on Google Play.\footnote{{https://play.google.com/store/apps/details?id=edu.northeastern.ccs.proxydetect}}
Users who downloaded and ran this app provided 23 sets of measurements (260 samples) from 11 networks. 
The study was conducted with IRB approval (NEU IRB\#15-04-07); users are consented before participation and absolutely 
no personal information is collected.

\vspace{-1em}
\subsection{Web Proxy Detection Methodology}
A key challenge for our work is detecting proxies \emph{without access to low-level packet traces, nor access to Web servers}. 
Our methodology must therefore provide a way to infer that there is a Web proxy based only at information gleaned from 
the application layer at the client.  Achieving this goal requires us to address two problems: 1) controlling when traffic 
is subject to proxy interposition, and 2) detecting the impact of this interposition. 

To address the first problem, we rely on results from previous studies (and reconfirmed in our own study), that Web 
proxies operate on traffic only to certain ports (\eg, port 80 for HTTP), but not others (\eg, port 443 for HTTPS). We 
thus run back-to-back experiments to fetch the same Web page over HTTP and HTTPS from the same server,\footnote{As such, our methodology is limited to sites that support both protocols, which is common among top Web sites.} where the latter is not 
subject to proxy interposition. To address the second problem, we use a combination of features that include 
latency differences and content modification. In the following subsections, we provide details about each test.

\vspace{-1em}
\subsubsection{Proxy Existence Test.}

%
%
%
\begin{figure}[tb]
\centering
\includegraphics[width=1\columnwidth]{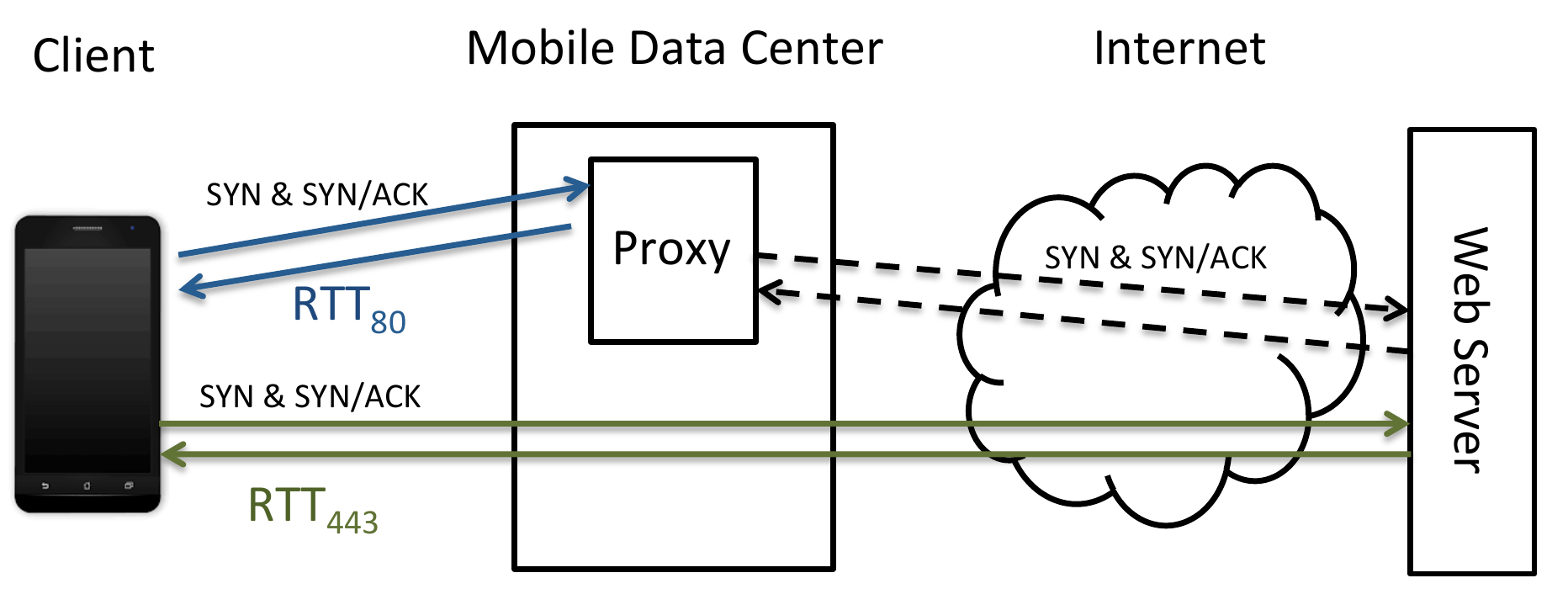}
\caption{{Diagram depicting a typical Web proxy that splits TCP connections on port 80 (HTTP) but not on port 443 (HTTPS). The Web proxy sends a SYN/ACK immediately in response to a client's SYN on port 80, but does not do so on port 443. Thus, $RTT_{80}$ should be less than $RTT_{443}$ if a Web proxy is present. }} 
\vspace{\postfigspace}

\label{fig:Proxy Detection.}
\end{figure}
We use a latency-based approach to infer the existence of Web proxies. Specifically, for each destination we create a TCP socket, and measure round-trip time (RTT) between the SYN and SYN/ACK as measured by the client. We do this on port 80 and port 443 in back-to-back measurements, then repeat this pair of measurements multiple times to account for noise in cellular network performance.\footnote{We also fix the IP address of the destination Web server to ensure that we contact the same host on both ports.} Based on our observations, a Web proxy will send a SYN/ACK to client's SYN immediately; however, a SYN packet sent to the same server on 443 will bypass the Web proxy and the reply will come directly from the destination server. 

We thus compare $RTT_{80}$ with $RTT_{443}$. We assume that the Web proxy is located somewhere on the path between client and Web server, and so if a Web proxy exists in the network, then $RTT_{80}$ should be less than $RTT_{443}$. Otherwise, the RTT measured on these two ports should be nearly identical.  

Because cellular network latency can vary significantly over time due to channel conditions and congestions, it is important to use multiple samples. However, there is a trade-off between increasing the number of samples and minimizing the data consumed by our tests for users running our experiments. To explore this issue, we investigated the impact of the number of tests on system accuracy (not shown). When varying the number of tests from 2 to 10 we found that 4 probes was the best number for accurately identifying proxies even in the presence of noise while minimizing the number of tests.

To understand the impact of proxies on popular Internet destinations, we selected Web sites from Alexa's top 100 global sites~\cite{topweb}. However, there are two challenges when selecting sites to use for identifying proxies. First, to achieve good QoE for Web users, many destination servers are already placed close to (or inside of) mobile networks, meaning the proxy and destination server paths are nearly identical. Using only such sites in our measurements can result in false negatives, since $RTT_{80}$ will be similar to $RTT_{443}$ (we explore this in Fig.~\ref{fig:detection-bad}). 

Second, noise in the cell network due to congestion, signal strength, or other factors can impact our inference by causing latency to vary significantly between tests. To reliably identify proxy behavior, we thus need to select destinations that are relatively ``far'' from the client, much more so than any variability in performance due to the network. 

To account for this noise, we propose the following dynamic approach that adjusts to the variance in any given measurement environment. We first measure   $RTT_{443}$ for the 100 top Web sites once. We then use the standard deviation (SD) of across RTT measurements, and use 2*SD as a threshold to exclude destinations that are ``too close'' to the network to be used for proxy inference. Specifically, for each destination, we compare $RTT_{443}$ with 2*SD, and see if the former is larger than the latter. We remove the websites whose $RTT_{443}$ is less than 2*SD.

When testing for a proxy, our client sets up TCP sockets with target Web servers on port 80 and port 443.  We use 4 probes for each test, \ie, for each destination we gather four pairs of $RTT_{80}$ and $RTT_{443}$. For each pair, we compute the difference between $RTT_{443}$ and $RTT_{80}$. After that, we compute the standard deviation (SD) of these four difference values. For each site, compare the average difference and the SD. If the difference value is greater than zero and is greater than SD, we infer that is a Web proxy in the network. 

We validated this approach for all US carriers in our study, using tests against a controlled server located in Amazon EC2 in Virginia. We found that our inference technique yielded identical results.

\vspace{-1em}
\subsubsection{Caching.}

To detect caching at Web proxies, we make the assumption that the fetch time for content from cache is smaller than from the origin server, based on the assumption that the cache is closer to the client in terms of latency. We thus use a latency-based methodology to infer whether there is a caching proxy in the network, identifying caching based on differences in fetch times when repeatedly fetching cacheable content. 

Specifically, we send two consecutive HTTP GETs to fetch the same (cacheable) object twice. Assuming an empty cache at the start of the experiment, and that the Web proxy caches the fetched content, the fetched object will be cached after the first HTTP GET. In this case, the second HTTP GET will be served from cache. We then compute the difference in download times for the two GETs ($first_{80}$ and $second_{80}$) and infer that an object was cached if $first_{80}$ is statistically significantly larger than $second_{80}$. 

Because a given object may have been cached by a previous fetch, we use multiple sources of content (some of which should not be cached) and ensure that our experiments leave sufficient time between pairs of GETs to ensure that any cached content expires.  We validated this approach with our controlled server, confirming that each entries expire. For the two carriers that deployed Web caches (Sprint and Boost), we found that cache entries will expire after about 5 minutes.

We dynamically determine which objects to fetch to test for caching by downloading the index page for popular Web sites and extracting URLs for embedded objects. 
We found that Web caches do not cache all file types. To explore this behavior, we include CSS, JavaScript, JPG, PNG, GIF and HTML files in our experiment because they are commonly embedded in Web pages.

Similar to the proxy detection scenario, we may not detect a proxy's caching behavior if the cache and origin server are relatively near each other and the file sizes are small. Thus, we reuse the ``far'' websites which we include in proxy detection experiment for URL extraction, and we exclude small objects (less than 5 KB).

\vspace{-1em}
\subsubsection{Object Rewriting.}
To detect object rewriting, we rely on the fact that a Web proxy cannot modify HTTPS traffic without breaking encryption, but is free to do so with HTTP traffic. We thus compare the results of fetching the same Web object over HTTPS and HTTP to determine if content modification on plaintext traffic. 

We use a similar approach to the cache-detection methodology for identifying objects to fetch, using embedded objects on popular Web sites \emph{and ensuring that servers hosting the content support both HTTP and HTTPS}. We fetch each object over HTTP (port 80) and HTTPS (port 443), and compare the contents of the fetched objects to identify rewriting. 

It is, of course, possible that Web content fetched from port 80 and 443 is not identical, and thus differences in content are not necessarily due to rewriting. To exclude this false positive, we use a server we control to verify wether the Web content differs when fetched using HTTP and HTTPS, based on the assumption that such behavior will not change based on the client's location. If the content is the same over both protocols for our control server, but they are different when fetched over a cellular network, we conclude that a Web proxy is rewriting HTTP content. 

We include CSS, JavaScript, PNG, GIF, JPG, and HTML files in the experiment to explore what file types are rewritten and how. Our approach allows us to detect behaviors such as transcoding, compression, header modification, and content injection.

\vspace{-1em}
\subsubsection{Redirection.}

Xu et al.~\cite{xu-transparent-proxies} showed that some Web proxies ignore the IP address provided HTTP GET requests, and instead perform a new DNS lookup to determine the destination IP based on resolving domain name in Host: field of HTTP GET request header. In this case, an HTTP flow may be sent to a different server than specified by the client. To detect such redirection behavior, we were unable to use latency-based techniques effectively so we use two servers under our control, E1 and E2, each having a different IP address. We modify the HTTP GET header content, putting the domain name of E1 in the {\tt Host:} field and sending the GET request to E2 (providing the IP address of E2 in the IP header). 

We do this both for E1 and E2, and check which server receives the HTTP request. If E1 receives the GET request, it indicates that a Web proxy redirected the GET request to E2 by resolving domain name of E1 into the IP address of E1. Otherwise, the proxy does not do redirection.
\vspace{-1em}
\section{Results}
In this section, we present the results of applying our methodology to the dataset described in the previous section. 
We summarize the results for proxy feature detection in Table~\ref{tab:summaryresults}. 

When compared with findings from a study conducted last year~\cite{xu-transparent-proxies}, we found that some carriers 
either changed their Web proxy deployments, and/or deploy different proxy features in different locations. For example, 
Xu et al. found that T-Mobile deployed a Web proxy that supported caching and traffic redirection. However, our results show there is no Web proxy at all anymore. 

By operating on unprivileged devices, our approach allows us to measure a wide range of networks worldwide that are difficult to measure if users must root their phones to conduct measurements~\cite{xu-transparent-proxies}. We found that Austrian, Chinese, and Canadian carriers do not deploy Web proxies. We will discuss our results in detail in the following subsections. 

  \begin{figure}
 
\mbox{
\subfigure[AT\&T and Boost]{
 \epsfig{file=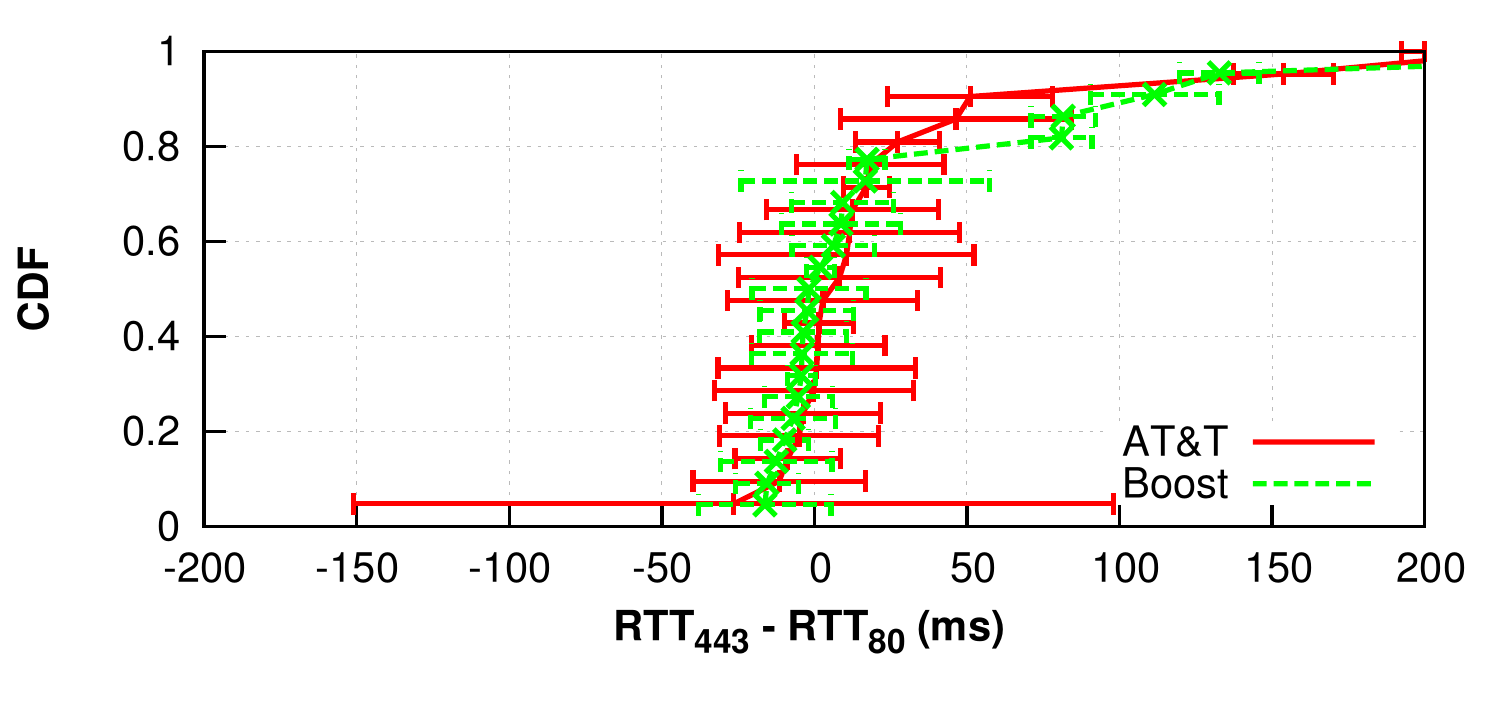, width=0.23\linewidth, angle=270}
 \label{fig:attboost}
}
\subfigure[Verizon and Sprint]{
 \epsfig{file=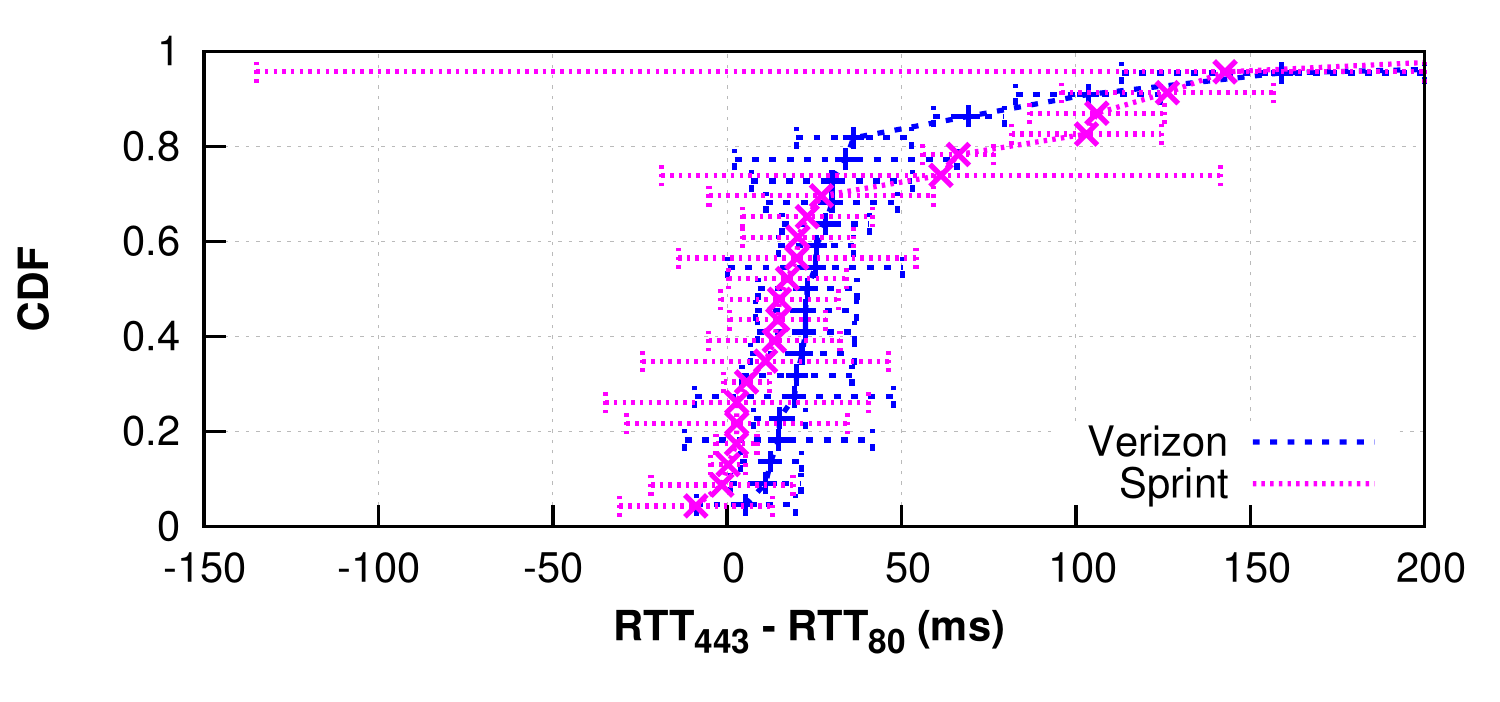, width=0.23\textwidth, angle=270}
 \label{fig:vzsprint}
}
}
\vspace{-0.5em}
\caption{\textbf{Proxy detection without filtering.} CDF of RTT differences between port 80 and port 443 on carriers with confirmed proxies, when using top Alexa Web sites. For most destinations, the difference between $RTT_{80}$ and $RTT_{433}$ is near zero because the servers are relatively close to the clients. Given the relatively large variance in latency (error bars are 1 SD), it is difficult to reliably infer proxies based solely on these results.}
\label{fig:detection-bad}
\vspace{-2.5em}
\end{figure}


\begin{figure}[tb]
\centering
\includegraphics[width=0.35\columnwidth, angle=270]{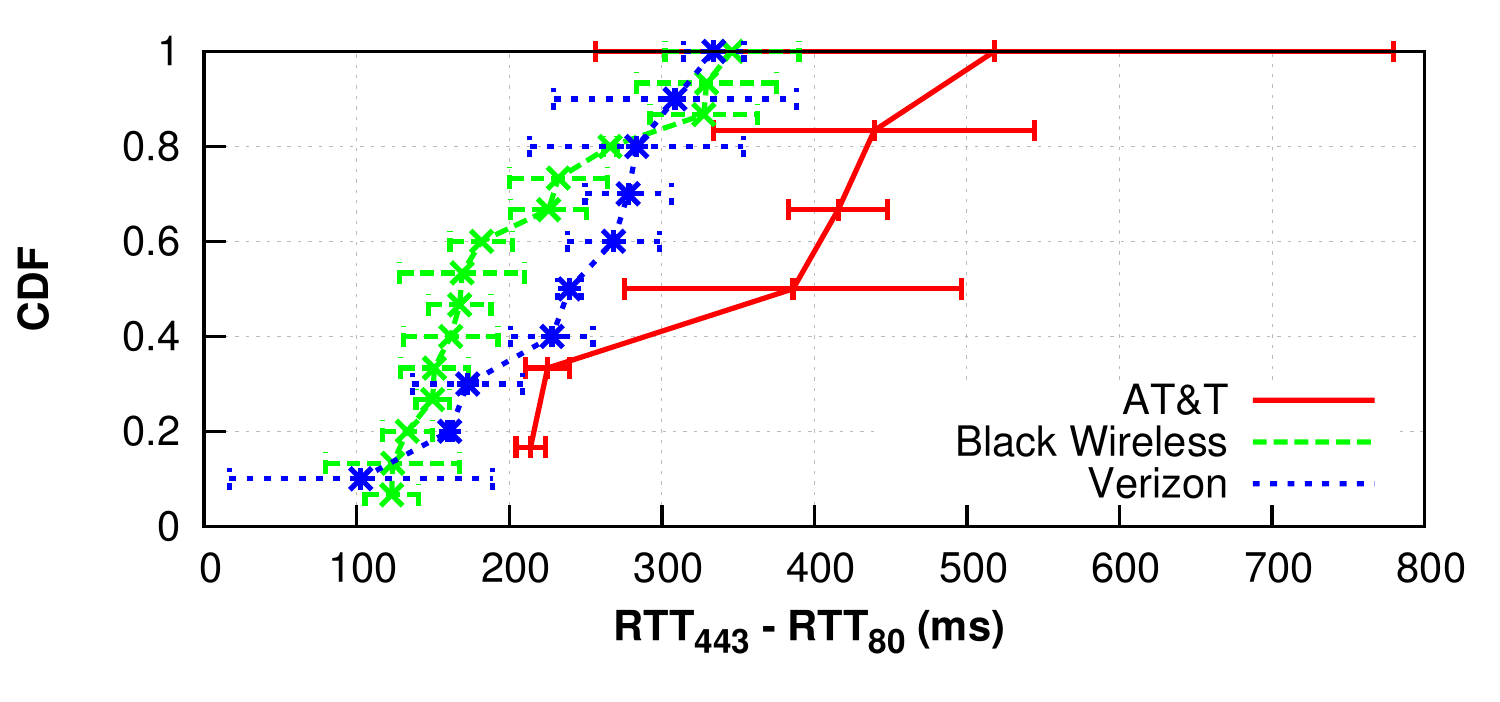}
\includegraphics[width=0.35\columnwidth, angle=270]{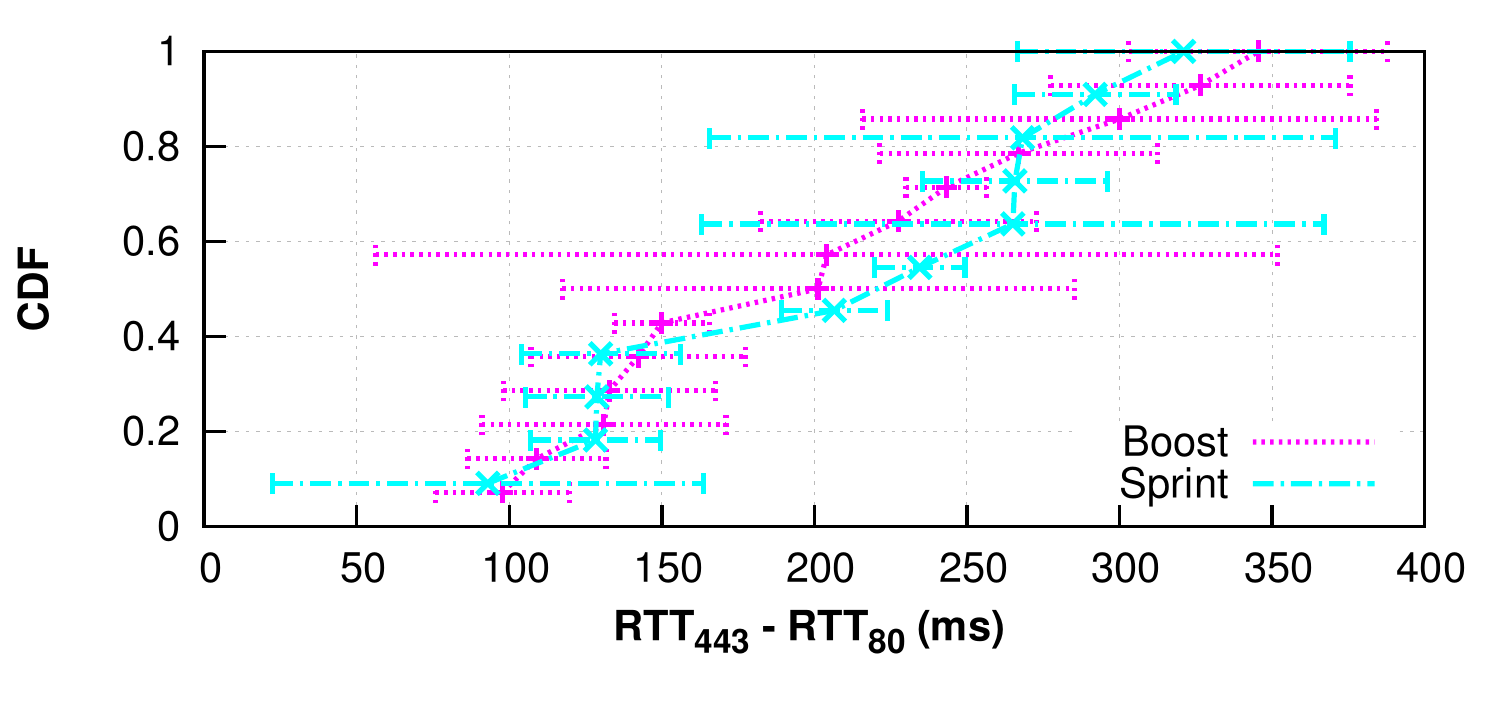}
\caption{\textbf{Proxy detection with filtering.} {CDF of RTT differences between port 80 and port 443, \emph{after filtering out nearby sites}. With filtering, $RTT_{443}$ is consistently larger than $RTT_{80}$ for all destinations, and the difference between the two is significantly larger than the noise. Our results indicate that AT\&T, Black Wireless, Verizon, Boost, and Sprint deploy Web proxies.}}
\vspace{\postfigspace}
\label{fig:detection-filter}
\end{figure}

\begin{figure}[tb]
\centering
\includegraphics[width=0.35\columnwidth, angle=270]{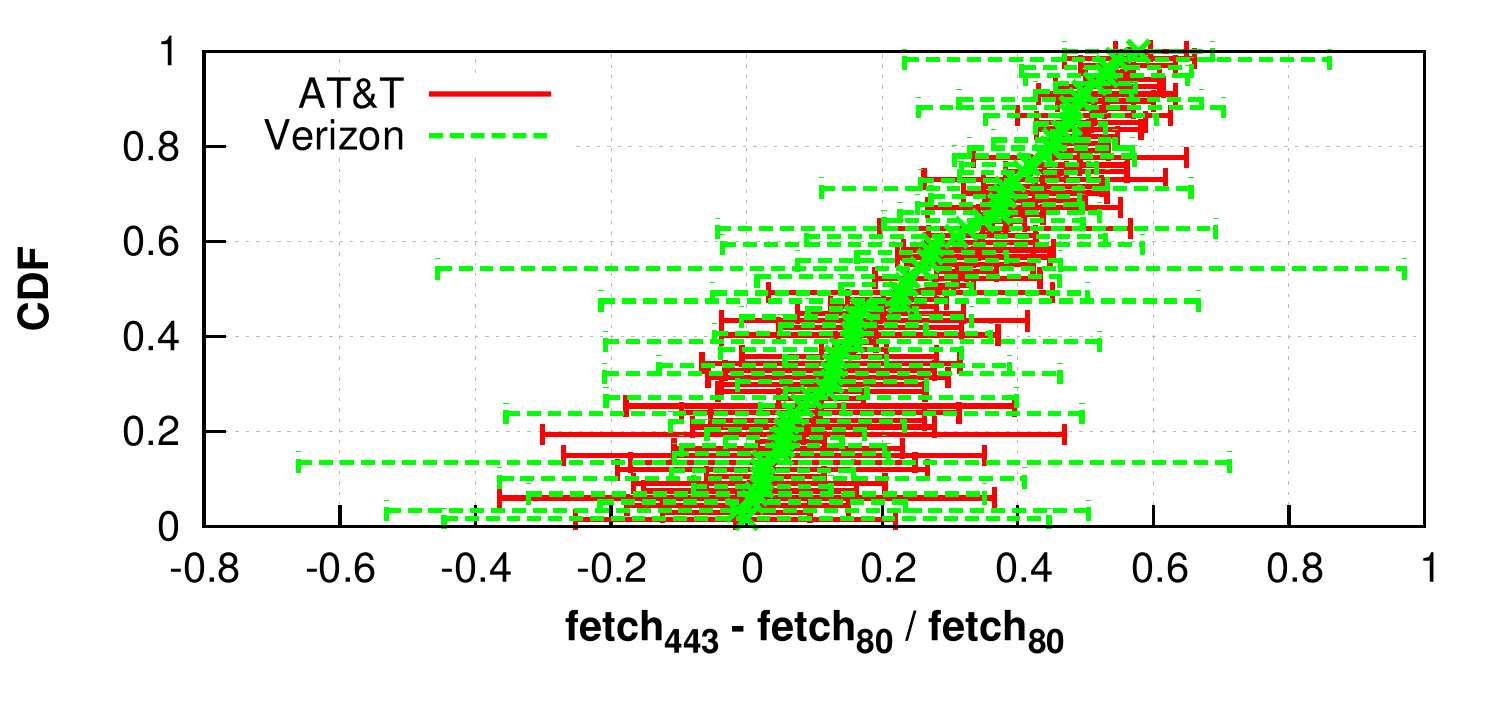}
\caption{\textbf{Impact of split TCP connections on HTTP GET performance.} The split TCP connections can improve end-to-end performance, though most performance gains are modest and not significantly large (64\% of AT\&T samples and 54\% of Verizon samples have an improvement that is less than 1 SD). For AT\&T, the maximum performance improvement is 60\%, and the maximum for Verizon is  57\%. }
\vspace{\postfigspace}
\label{fig:endtoend-performance}
\end{figure}

\vspace{-1em}
\subsection{Web Proxy Detection}

We identified Web proxies in AT\&T, Verizon, Boost, Sprint, and Black Wireless. To demonstrate the effectiveness of our methodology, we evaluate what happens when we do not filter nearby Web sites as discussed in the previous section. In Fig.~\ref{fig:detection-bad}, we show the differences between RTTs on ports 80 and port 443 (using error bars representing the standard deviation) \emph{without filtering nearby sites} for carriers that we know use Web proxies. The figure clearly shows that the variance is much larger than the RTT differences for many of the cases, and that most of the RTT differences are near zero. Without filtering, it is difficult to draw strong conclusions about proxies being present in the network.

In Fig.~\ref{fig:detection-filter}, we show detection results when excluding ``nearby'' sites. In this case, the RTT differences are significantly greater than zero, and are larger than the standard deviation in the measurements, leading us to conclude strongly that a proxy is present. In some networks, the RTT difference may not exceed the noise level for all sites. To account for this, we conclude a proxy is present if the latency difference is greater than one standard deviation for 80\% of the sites.

In Fig.~\ref{fig:endtoend-performance}, we investigate how Web proxies impact end-to-end performance for HTTP. We fetch objects using port 80 and port 443, and compare object fetch times. We focus only on AT\&T and Verizon, because Sprint and Boost rewrite and cache objects, which may bias our results. Our results indicate that Web proxies can reduce object fetch time for popular pages, but the impact is often not statistically significant. For AT\&T, nearly two thirds of the samples have a SD that exceeds the performance improvement, and the same is true for 54\% of Verizon samples. Nonetheless, there are cases of significant performance improvement---up to 60\% in AT\&T and 57\% in Verizon.

\vspace{-1em}
\subsection{Caching}
\begin{figure}[tb]
\centering
\includegraphics[width=0.35\columnwidth, angle=270]{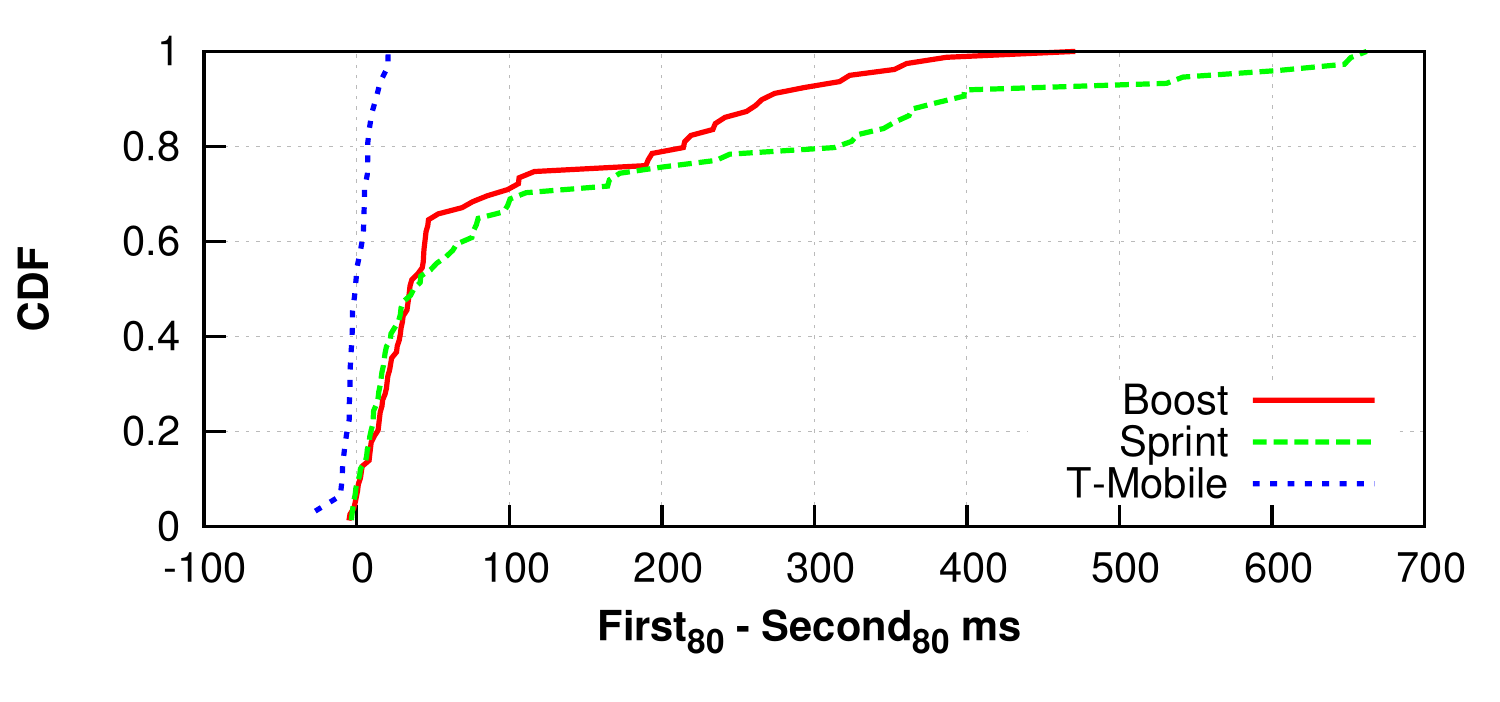}
\caption{\textbf{Cache detection.} {CDF of the difference between object fetch time for the first GET and second GET, for objects that are cacheable. Note that T-Mobile does not cache content, so most values are near zero. However, Boost and Sprint do cache content, and the second GET is noticeably shorter than the first. }
{\sl}}
\vspace{\postfigspace}
\label{fig:caching-}
\vspace{-1em}
\end{figure}

We detected caching in Sprint and Boost networks, as shown in Fig.~\ref{fig:caching-}. They cache static CSS, JavaScript, PNG, JPG, and GIF files, but do not cache HTML. We also conducted experiments to determine if the Web caches respect cache control headers (\eg setting {\tt cache-control: no-store}, and checking whether an object is cached in violation of the header). We find that Sprint and Boost respect cache control headers. 

\begin{figure}[tb]
\centering
\includegraphics[width=0.35\columnwidth, angle=270]{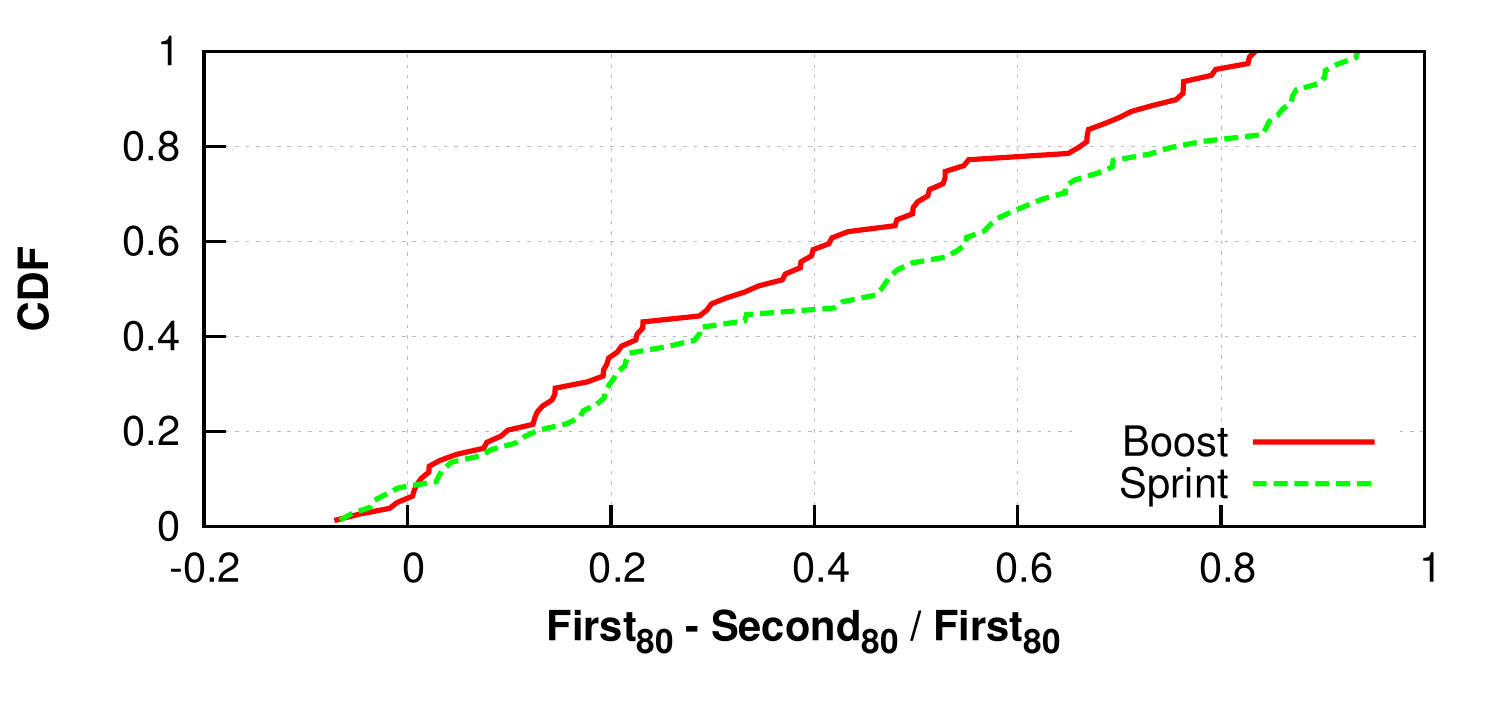}
\caption{CDF of relative performance gains from caching. The median difference in performance is 33\% for Boost and 46\% for Sprint.}
\vspace{\postfigspace}
\label{fig:caching-performance}
\end{figure}

We determined the impact of caching on object fetch time. In Fig.~\ref{fig:caching-performance}, we show the relative change in fetch time for cached content compared to content fetched from the origin. The median and maximum difference for Boost is 33\% and 83\%, respectively; for Sprint, the improvements are 46\% and 93\%.\footnote{For clarity, we omit the SD error bars from here out, but the impact of noise is similar to what we found in the proxy detection examples.}

\vspace{-1em}
\subsection{Object Rewriting}
\begin{figure}[tb]
\centering
\includegraphics[width=0.7\columnwidth]{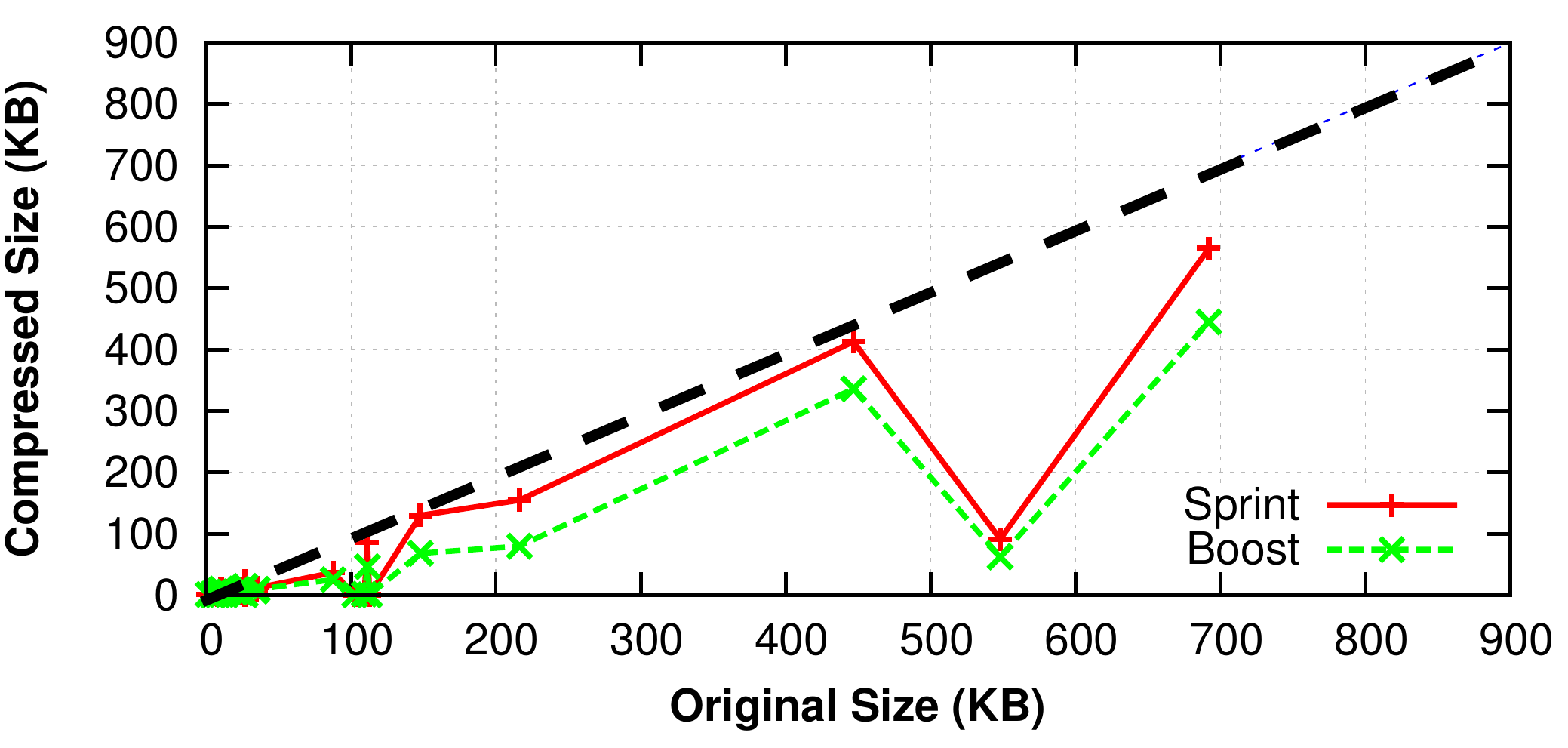}
\caption{\textbf{Content modification.} {Boost and Sprint  were found to transcode image files to lower quality, with Boost reducing image quality 
more aggressively than Sprint. Neither carrier transcodes images larger than $\approx$700KB. Dashed line indicates $y=x$.}}
\vspace{\postfigspace}
\label{fig:content-rewriting}
\end{figure}

Our analysis identified several cases of object rewriting. Specifically, Boost and Sprint compress image files (JPG, GIF and PNG). Black Wireless compresses CSS and JavaScript files. Interestingly, even though Boost is an MVNO running atop Sprint, the two carriers compress the same image file into different sizes. In particular, Boost is more aggressive in reducing file size (and thus image quality) compared with Sprint proxy, as Fig.~\ref{fig:content-rewriting}. 
We also observed that neither carrier modifies image files larger than 700\,KB, which is bigger than the file size threshold observed by Xu et al~\cite{xu-transparent-proxies}.

\begin{figure}[tb]
\centering
\includegraphics[width=0.35\columnwidth, angle=270]{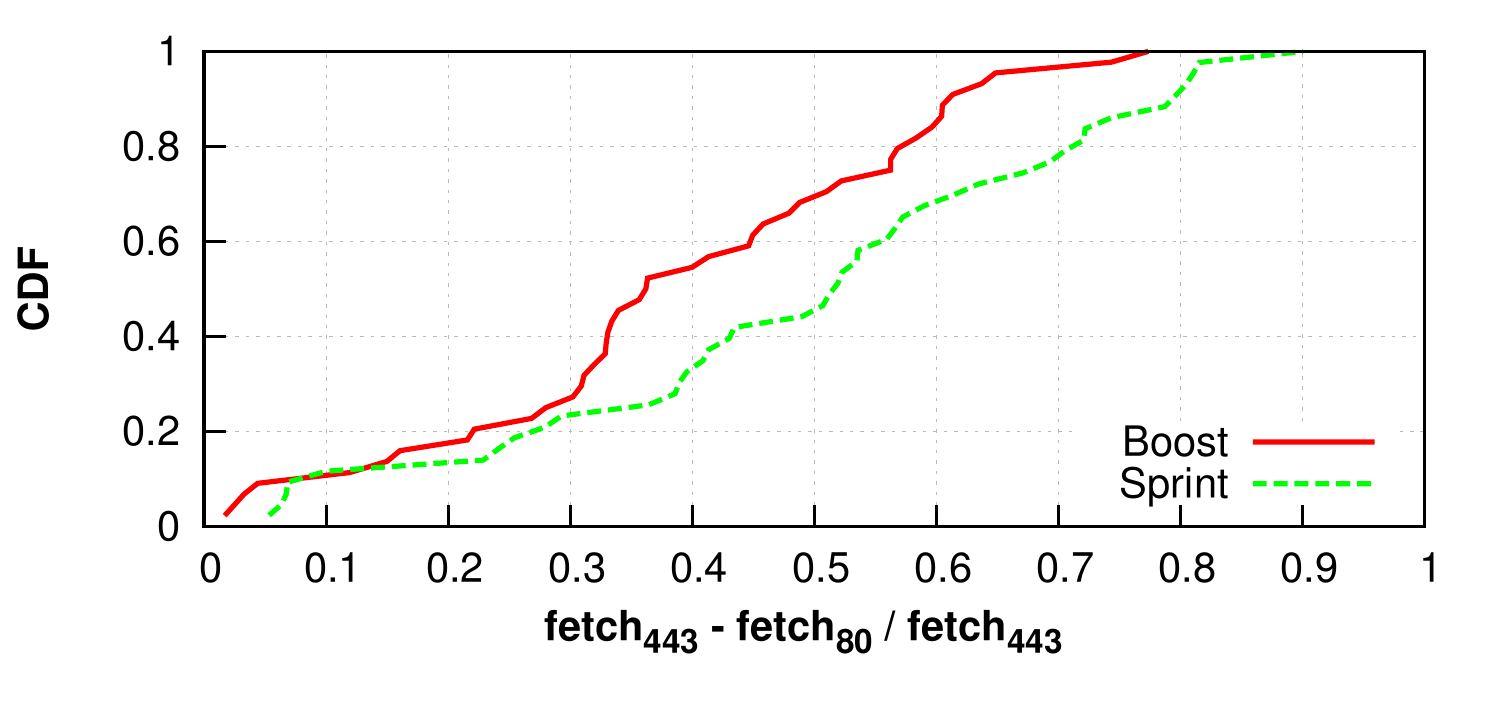}
\caption{Compressed objects can be fetched up to 77\% and 90\% faster than the original files for Boost and Sprint, respectively. 
The median performance gains are 36\% and 52\%. }
\vspace{-1em}
\label{fig:rewriting-performance}
\end{figure}

While image quality is reduced, such transcoding can certainly improve performance. In Fig.~\ref{fig:rewriting-performance}, we quantify the relative improvement in fetch times for transcoded vs non-transcoded images. Boost sees up to 77\% improvement, and Sprint sees up to  90\%. The median performance enhancement is 36\% for Boost, and 52\% for Sprint. It is interesting to note that while Boost compresses images more aggressively, their improvement in fetch times is lower than Sprint. While we were not able to isolate the root cause, it is possible that more aggressive transcoding is slowing down the file transfer. 

Interestingly, Black Wireless and AT\&T do not share the same rewriting behavior, even though Black Wireless runs atop AT\&T. Black Wireless deploys a compression proxy that rewrites certain text content, but AT\&T does not modify any content type in our experiments.

\vspace{-1em}
\subsection{Redirection}
We used our methodology to check whether any of the carriers redirect traffic by conducting additional DNS lookups. We found no instances of this behavior. Note that Xu et al.~\cite{xu-transparent-proxies} showed that T-Mobile's Web proxy performed redirection one year prior to this study, and this behavior has since changed.

\vspace{-1em}
\section{Conclusion}
\vspace{-1em}
This paper described a technique for detecting Web proxy features from mobile devices 
without requiring privileged access at clients, nor control over the Web servers being 
measured. This approach allowed us to characterize proxies from 11 networks worldwide, 
and to understand their impact on end-to-end performance. We find that proxy features 
change over time, and that their impact can be significant on HTTP fetch times and quality of content 
for some cases. We believe that our work opens new opportunities for improving 
transparency and performance in mobile networks by revealing policies through 
crowdsourced measurements. As part of our future work, we will integrate our 
approach into Mobilyzer, implement an iOS version of the app, and provide a public 
Web site where users, operators, and policymakers can view carriers' middlebox 
deployments and their impact on traffic.

\vspace{-1em}

\bibliographystyle{abbrv}
\bibliography{mobile-platform-unified}

\end{document}